\documentclass[a4paper,12pt]{article}
\usepackage{epsfig,float}
\usepackage{amsmath,amsfonts,amssymb}
\usepackage[left=2cm,right=2cm]{geometry}
\usepackage{a4wide,graphicx}

\newcommand{\be}{\begin{equation}}
\newcommand{\ee}{\end{equation}}
\newcommand{\ba}{\begin{eqnarray}}
\newcommand{\ea}{\end{eqnarray}}

\newcommand{\De}{\Delta}

\newcommand{\Sg}{\Sigma^*}

\newcommand{\X}{\Xi^*}

\newcommand{\Om}{\Omega}

\begin{document}

\title{Radiative decay into $\gamma$-baryon of dynamically generated resonances 
from the vector-baryon interaction}

\author{Bao-Xi Sun$^{1}$, E.J. Garzon$^2$, and E. Oset$^2$}
\maketitle

\begin{center}
$^1$ Institute of Theoretical Physics, College of Applied Sciences,\\
Beijing University of Technology, Beijing 100124, China\\
$^2$ Departamento de F\'{\i}sica Te\'orica and IFIC,
Centro Mixto Universidad de Valencia-CSIC,
Institutos de Investigaci\'on de Paterna, Aptdo. 22085, 46071 Valencia, Spain \\
\end{center}

\date{}

 \begin{abstract}
 We study the radiative decay into $\gamma$ and a baryon of the SU(3) octet and decuplet
 of nine and ten resonances that are dynamically generated from
 the interaction of vector mesons with baryons of the octet and the decuplet respectively. We obtain
 quite different partial decay widths for the various resonances, and for
 different charge states
 of the same resonance, suggesting that the experimental investigation
 of these radiative decays should bring much information on the nature of these
 resonances.
 For the case of baryons of the octet we determine the helicity amplitudes and compare 
 them with experimental data when available.
\end{abstract}

\section{Introduction}

  In a recent paper \cite{pedro}, the $\rho  \Delta$ interaction was studied
  within the local hidden gauge formalism for the interaction of vector mesons.
  The results of the interaction gave a natural
  interpretation  for the $\Delta(1930)(5/2^-)$ as a  $\rho  \Delta$
  bound state,  which otherwise is extremely problematic in quark models since
  it involves a $3  h \omega$ excitation and appears with much higher mass.
   At the same time two states with $J^P=1/2^-, 3/2^-$ were
  obtained, degenerate with the  $5/2^-$, which could be accommodated with
  two known $\Delta$ states in that energy range.  Also, three 
  degenerate $N^*$ states with $1/2^-, 3/2^-,5/2^- $ were obtained, which were more
  difficult to identify with known resonances since that sector is not so well
  established.  The work of \cite{pedro} was extended to the SU(3) sector in
  \cite{sourav} to
  account for the interaction of vectors of the octet with baryons of the
  decuplet. In this case ten resonances, all of them also degenerate in the three
  spin states, were obtained, many of which could be identified with existing
  resonances, while there were predictions for a few more.   At the same time
  in \cite{sourav} the poles and residues at the poles of the resonances were
  evaluated, providing the coupling of the resonances to the different
  vector-baryon of the decuplet components.

  One of the straightforward tests of these theoretical
  predictions is the radiative decay of these resonances into photon and the
  member of the baryon decuplet to which it couples. Radiative decay of
  resonances into $\gamma N$ is one of the observables traditionally
   calculated 
  in hadronic models. Work in quark models on this issue is abundant
\cite{Darewych:1983yw,Warns:1990xi,Umino:1991dk,Umino:1992hi,Bijker:2000gq,
close,Konen:1989jp,Warns:1989ie,capstick,capstickcont,Aiello:1998xq,Pace:1998pp,
metsch,santopinto,VanCauteren:2003hn,VanCauteren:2005sm}.  For resonances which appear as dynamically generated in
chiral unitary theories there is also much work done on the radiative decay
into $\gamma N$ \cite{kaiser,Borasoy:2002mt,mishasourav,mishasolo,mishageng}.
Experimental work in this topic is also of current interest
 \cite{Thompson:2000by,burkert,Aznauryan:2004jd}.

 In the present work we address the novel aspect of radiative decay into a
 photon and a baryon of the decuplet of the $\Delta$, since the underlying
 dynamics of the resonances that we study provides this  as
 the dominant mode of radiative decay into photon baryon.
  This is so, because the underlying
  theory of the studies of \cite{pedro,sourav} is the local hidden gauge
  formalism for the interaction of vector mesons developed in
  \cite{hidden1,hidden2,hidden3,hidden4}, which has the peculiar feature, inherent to vector
  meson dominance, that the photons couple to the hadrons through the
  conversion into a vector meson.  In this
  case a photon in the final state comes from either a $\rho^0, \omega, \phi$.
  Thus, the radiative decay of the resonances into $\gamma B$ is readily obtained
  from the theory by taking the terms with $\rho^0 B, \omega B, \phi B $ in the
  final state and coupling the $\gamma $ to any of the final
  $\rho^0, \omega, \phi  $ vector mesons.  This procedure was used in
  \cite{junko} and provided
  good results for the radiative decay into $\gamma \gamma$ of the $f_0(1370)$
  and $f_2(1270)$ mesons which were dynamically generated from the $\rho
  \rho$ interaction within the same framework  \cite{raquel}. This latter work
  was also extended to the interaction of vectors with themselves within
  SU(3), where many other states are obtained which can be also associated with
  known resonances \cite{geng}.
  The radiative decay of the latter resonances into $\gamma \gamma$ or a $\gamma$ and
  a vector has been studied in \cite{Branz:2009cv}, with good agreement with experiment
  when available.
  Given the success of the theory in its
  predictions and the good results obtained for the
  $\gamma \gamma$ decay of the $f_0(1370)$, $f_2(1270)$ and $f'_2(1525)$ mesons, the
  theoretical framework stands on good foot and the predictions made should be
  solid enough to constitute a test of the theory by contrasting with experimental
  data.
  The extension of the work of \cite{pedro,sourav} to the interaction of vector mesons with baryons
  of the octet of the proton has also been successful \cite{angelsvec} and nine resonances,
  degenerated in spin-parity $1/2^-$ and $3/2^-$, appear dynamically generated in the
  approach, many of which can be naturally associated to know resonances in the PDG \cite{pdg}.
  We also extend the present work to study the radiative decay of these resonances into
  a photon and a baryon of the octet. In this case we can also evaluate helicity amplitudes
  and compare them with experimental results when available.

       The  experimental situation in that region of energies is still poor.
  The PDG \cite{pdg} quotes many radiative decays of $N^*$ resonances, and of the
  $A_{1/2},A_{3/2}$ helicity amplitudes for decay of resonances into $\gamma N$,
  with N either proton or neutron.
  However, there are no data to our knowledge for radiative decay into
  $\gamma B$, with B a baryon of the decuplet. The reason for it might be the
  difficulty in the measurement, or the lack of motivation, since there are also
  no theoretical works devoted to the subject.  With the present work we hope to
  reverse the situation offering a clear motivation for these experiments since
  they bear close connection with the nature invoked for these resonances,
   very different to the ordinary three quark structure of the baryons.
   
   The numbers obtained for the radiative widths are well within measurable
   range, of the order of 1 MeV, and the predictions are interesting, with
   striking differences of one order of magnitude between decay
   widths for different charges of the same resonance.

    The work will proceed as follows. In the next two Sections we present the
    framework for the evaluation of amplitudes of radiative decay.  In Section 4
	we show the results obtained for the different resonances generated with the
	baryon decuplet. Section 5 introduces the equations for the baryon octet, which
	are used in Section 6 to obtain results for the deacy width of the
	resonances dynamically generated with a vector and the baryon octet.
	In Section 7 we present the results for the helicity amplitudes of some
	resonances used in the previous section, and in Section 8 we finish with some conclusions.

\section{Framework}

\begin{figure}[htb]
\begin{center}
\includegraphics[width=0.5\textwidth]{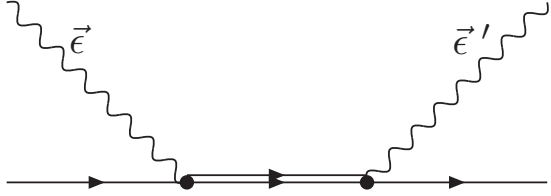}
\end{center}
\caption{Diagram contributing to the vector-baryon interaction via
the exchange of a resonance.} \label{fig:feyn1}
\end{figure}

\begin{figure}[htb]
\begin{center}
\includegraphics[width=0.35\textwidth]{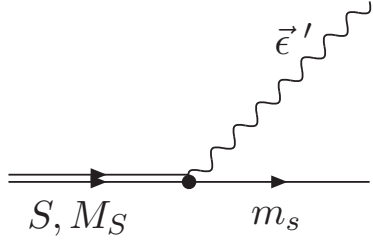}
\end{center}
\caption{Diagram on the decay of the resonance in a decuplet baryon
and a vector meson.}  \label{fig:feyn2}
\end{figure}

In Ref.~\cite{pedro,sourav}, the scattering amplitudes for
vector-decuplet baryon $VB\rightarrow V^\prime B^\prime$ are
given by
\begin{equation}
\label{eq:tvbvb}
 t_{VB\rightarrow V^\prime B^\prime}~=~t~\vec{\epsilon} \cdot
\vec{\epsilon^\prime}~ \delta_{m_s,m^\prime_s},
\end{equation}
where $\vec{\epsilon}$, $\vec{\epsilon\prime}$ refer to the initial
and final vector polarization and the matrix is diagonal in the
third component of the baryons of the decuplet. The transition is
diagonal in spin of the baryon and spin of the vector, and as a
consequence in the total spin. To make this property more explicit,
we write the states of total spin as
\begin{equation}
|S,M\rangle~=~\sum_{m_s}C \left(3/2, 1, S; m_s, M-m_s,
M\right)|3/2,m_s\rangle |\vec{\epsilon}_{M-m_s}\rangle
\end{equation}
and
\begin{equation}
 \langle S,M|~=~\sum_{m^\prime_s}C \left(3/2, 1, S; m^\prime_s, M-m^\prime_s,
M\right)\langle 3/2,m^\prime_s|  \langle
\vec{\epsilon}_{M-m^\prime_s}^{~*}|,
\end{equation}
where $C \left(3/2, 1, S; m_s, M-m_s, M\right)$  are the Clebsch-Gordan
coefficients and $\epsilon_{\mu}$ the polarization vectors in
spherical basis
\begin{equation}
\vec{\epsilon}_+~=~-\frac{1}{\sqrt{2}}\left(\vec{\epsilon}_1~+~i\vec{\epsilon}_2\right),~~~~
\vec{\epsilon}_-~=~\frac{1}{\sqrt{2}}\left(\vec{\epsilon}_1~-~i\vec{\epsilon}_2\right),~~~~
\vec{\epsilon}_0~=~\vec{\epsilon}_3.
\end{equation}
We can write Eq.~(\ref{eq:tvbvb}) in terms of the projectors
$|S,M\rangle \langle S,M|$ as
\begin{equation}
\label{eq:tvbvb2}
 t_{VB\rightarrow V^\prime B^\prime}~=~t~\langle \vec{\epsilon^\prime}|\langle
 3/2, m^\prime_s | \sum_{S, M} |S, M\rangle \langle S, M| 3/2, m_s
 \rangle | \vec{\epsilon}~ \rangle.
\end{equation}
Since the Clebsch-Gordan coefficients satisfy the normalization
condition
\begin{equation}
\sum_{S}C \left(3/2, 1, S; m_s, M-m_s, M\right)C \left(3/2, 1, S;
m^\prime_s, M^\prime-m^\prime_s,
M^\prime\right)~=~\delta_{m_sm^\prime_s}\delta_{MM^\prime},
\end{equation}
we have then
\begin{eqnarray}
\sum_{S, M} |S, M\rangle \langle S, M|&=&\sum_M \sum_{m_s}|3/2,
m_s\rangle\langle 3/2, m_s|~| \vec{\epsilon}_{M-m_s}~ \rangle\langle
\vec{\epsilon^*}_{M-m_s}| \\ \nonumber
&=&\sum_{M^\prime}
\sum_{m_s}|3/2, m_s\rangle\langle 3/2, m_s|~|
\vec{\epsilon}_{M^\prime}~ \rangle\langle
\vec{\epsilon^*}_{M^\prime}|\equiv1.
\end{eqnarray}
We can depict the contribution of a specific resonant state of spin
$S$ to the amplitude described by means of Fig.~\ref{fig:feyn1}.
Then the amplitude for the transition of the resonance to a final
vector-baryon state is depicted by means of Fig.~\ref{fig:feyn2}.
As shown is Ref.~\cite{pedro,sourav}, the $VB \to V^\prime B^\prime$
scattering amplitudes develop poles corresponding to resonances and
a resonant amplitude is written as Eq.~(\ref{eq:tvbvb}) with $t$
given by
\begin{equation}
t_{ij}~=~\frac{g_i g_j}{\sqrt{s}-M+i\Gamma/2}
\end{equation}
with $g_i$ and $g_j$ the couplings to the initial and final states.
Accordingly, the amplitude for the transition from the resonance to
a final state of vector-baryon is given by
\begin{eqnarray}
\label{eq:tsmvb}
 t_{SM\rightarrow V^\prime B^\prime}
 &=&g_i\langle\vec{\epsilon}|
 \langle 3/2, m_s | S, M\rangle \nonumber \\
 &=&g_i C(3/2, 1, S; m_s, M-m_s, M)\langle
 \vec{\epsilon}~|\vec{\epsilon}_{M-m_s}\rangle.
\end{eqnarray}
When calculating the decay width of the resonance into $VB$ we will
sum $|t|^2$ over the vector and baryon polarization, and average
over the resonance polarization $M$. Thus, we have
\begin{eqnarray}
\label{eq:sumvector}
 &&\frac{1}{2S+1}\sum_{M, m_s, \vec{\epsilon}}
|t_{SM\rightarrow V^\prime B^\prime}|^2 \\ \nonumber
&=&|g_i|^2
\frac{1}{2S+1}\sum_{M, m_s, \vec{\epsilon}}C(3/2,1,S;m_s,M-m_s,M)^2
\langle \vec{\epsilon^*}_{M-m_s}
 | \vec{\epsilon} \rangle
\langle \vec{\epsilon}|\vec{\epsilon}_{M-m_s}\rangle \\ \nonumber
&=&
|g_i|^2 \frac{1}{2S+1}\sum_{M^\prime} \sum_{m_s} \frac{2S+1}{3}
C(3/2,S,1;m_s,m_s+M^\prime,-M^\prime)^2
 \langle \vec{\epsilon^*}_{M^\prime} |
\vec{\epsilon}_{M^\prime} \rangle \\ \nonumber
&=& |g_i|^2
\frac{1}{3} \sum_{M^\prime} \delta_{M^\prime M^\prime }
\\ \nonumber
&=&|g_i|^2,
\end{eqnarray}
where in the first step we have permuted the two last spins in the
Clebsch-Gordan coefficients and in the second we applied their
orthogonality condition.

We observe that the normalization of the amplitudes is done in a way
such that the sum and average of $|t|^2$ is simply  the modulus
squared of the coupling of the resonance to the final state. The
width of the resonance for decay into $VB$ is given by
\begin{equation}
\label{eq:vectorwidth} \Gamma~=~\frac{M_B}{2\pi M_R}~ q~ |g_i|^2,
\end{equation}
where $q$ is the momentum of the vector in the resonance rest frame
and $M_B$, $M_R$ the masses of the baryon and the resonance. We
should note already that later on when the vector polarizations are
substituted by the photon polarizations in the sum over $M^\prime$
in Eq.~(\ref{eq:sumvector}) we will get a factor two rather than
three, because we only have two transverse polarizations, and then
Eq.~(\ref{eq:vectorwidth}) must be multiplied by the factor $2/3$.

\section{Radiative decay}

Next we study the radiative decay into $B \gamma$ of the resonances
dynamically generated in Ref.~\cite{sourav} with $B$ a baryon of the
decuplet. Recalling the results of~\cite{sourav} we obtained there
ten resonances dynamically generated, each of them degenerated in
three states of spin, $1/2^-$, $3/2^-$, $5/2^-$. As we have
discussed in the former section, the radiative width will not depend on the
spin of the resonance, but only on the coupling which is the same
for all three spin states due to the degeneracy. This would be of
course an interesting experimental test of the nature of these
resonances.

\begin{figure}[htb]
\begin{center}
\includegraphics[width=0.35\textwidth]{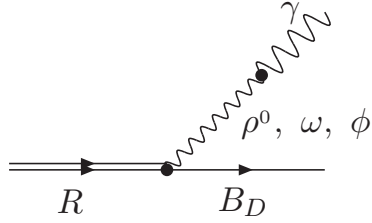}
\end{center}
\caption{Diagram on the radiative decay of the resonance in a
decuplet baryon and a photon.} \label{fig:feyn3}
\end{figure}

In order to proceed further, we use the same formalism of the hidden
gauge local symmetry for the vector mesons
of~\cite{hidden1,hidden2,hidden3,hidden4}. The peculiarity of this theory concerning
photons is that they couple to hadrons by converting first into a
vector meson, $\rho^0$, $\omega$, $\phi$. Diagrammatically this is
depicted in Fig.~\ref{fig:feyn3}. This idea has already been applied
with success to obtain the radiative decay of the $f_2(1270)$,
$f_0(1370)$, $f'_2(1525)$ and $f_0(1710)$ resonances into $\gamma \gamma$
in Ref.~\cite{junko,Branz:2009cv}. In Ref.~\cite{junko} the question of 
gauge invariance was addressed and it was
shown that the theory fulfills it. In
Ref.~\cite{hideko}, it is also proved in the case of radiative decay
of axial vector resonances.

The amplitude of Fig.~\ref{fig:feyn3} requires the $ \gamma V$
convertion Lagrangian, which comes from Refs.~\cite{hidden1,
hidden2,hidden3} and is given by $(see~ Ref.~\cite{hideko}~ for~ practical
~details)$
\begin{equation}
\label{eq:Lvgamma} {\cal L}_{V\gamma}~=~-M_V^2\frac{e}{\tilde{g}}
A_\mu\langle V^\mu Q\rangle
\end{equation}
with $A_\mu$ the photon field, $V_\mu$ the SU(3) matrix of vector
fields
 \be
  V_\mu \equiv  \left(\begin{array}{ccc}
\frac{1}{\sqrt{2}} \rho^0 + \frac{1}{\sqrt{2}}\omega
 & \rho^+ & K^{*+}\\
\rho^-& - \frac{1}{\sqrt{2}} \rho^0 + \frac{1}{\sqrt{2}}\omega
& K^{*0}\\
K^{*-}& \bar{K}^{*0} & \phi
\end{array}
\right)_{\mu}, \label{eq:PVmatrices} \ee
and $Q$ the charge matrix
\be Q\equiv \left(
\begin{array}{ccc}
2/3 & 0 & 0\\
0& -1/3 & 0\\
0&    0 & -1/3
\end{array}
\right).
 \ee
In Eq.~(\ref{eq:Lvgamma}), $M_V$ is the vector meson mass, for which we take
an average value $M_V=800MeV$, $e$ the electron charge, $e^2=4\pi
\alpha $, and
$$\tilde{g}=\frac{M_V}{2f};~~~~f=93MeV.$$
The sum over polarizations in the intermediate vector meson, which
converts the polarization vector of the vector meson of the
$R\rightarrow B V$ amplitude into the photon polarization of the
$R\rightarrow B \gamma$ amplitude, leads to the equation
\begin{equation}
-it_{\gamma V} D_V~=~-i M_V^2~\frac{e}{\tilde{g}}~ \frac{i}{-M_V^2}~
F_j
\end{equation}
with \be F_j~=~\{
\begin{array}{ccc}
\frac{1}{\sqrt{2}} & for &\rho^0, \\
\frac{1}{3\sqrt{2}}& for &\omega, \\
-\frac{1}{3}       & for &\phi.
\end{array}
 \ee
Thus, finally our amplitude for the $R\rightarrow B \gamma$ transition,
omitting the spin matrix element of Eq.~(\ref{eq:tsmvb}), is given
by

\be \label{eq:tgamma}t_\gamma~=~-\frac{e}{\tilde{g}}
\sum_{j=\rho^0,~\omega,~\phi}g_j F_j. \ee

 As discussed in the former section, the radiative decay
width will then be given by

\be
\label{eq:Ggamma}\Gamma_\gamma~=~\frac{1}{2\pi}~\frac{2}{3}~\frac{M_B}{M_R}~q~|t_\gamma|^2.
\ee

The couplings $g_j$ for different resonance and $VB$ with
$V=\rho^0,~\omega,~\phi$ and $B$ different baryon of the decuplet
can be found in Ref.~\cite{sourav} and we use them here for the
evaluation of $\Gamma_\gamma$. The factor $\frac{2}{3}$ in eq. (\ref{eq:Ggamma})
additional to eq. (\ref{eq:vectorwidth}) appears because now we have only two
photon polarizations and the sum over $M'$ in eq. (\ref{eq:sumvector}) gives 2
instead of 3 for the case of vector mesons.

\section{Results for radiative decays into $\gamma$ and baryon decuplet}

The couplings of the resonances to the different $VB$ channels are
given in Ref.~\cite{sourav} in the isospin basis. For the case of
$\omega B$ and $\phi B$, there is no change to be done, but for the
case of $\rho B$, one must project over the $\rho^0 B$ component.
Since this depends on the charge of the resonance $R$, the radiative
decays will depend on this charge, as we will see.
We recall that in our phase convention $|\rho^+~\rangle=~-
|1,1\rangle$ of isospin. The information on the resonances and their
couplings to different baryons of decuplet and vector mesons
$\rho$,$\omega$, $\phi$ for different channels is listed in
Table~\ref{tab:couple}. We detail the results below.

\begin{table}[hbt]
\begin{center}
\begin{tabular}{c|c|c|c|c}
\hline
S, I&Channel &  &  & \\
\hline
 & &  $z_{R}=1850+ i5$ & & \\
\hline
0, 1/2&$\De\rho$ & $4.9+i0.1$ &  &  \\
\hline
 & & $z_{R}=1972+i49$ & &  \\
\hline
0, 3/2&$\De\rho$ & $5.0+i0.2$ & & \\
      &$\De\omega$ & $-0.1+i0.2$ & & \\
      &$\De\phi$ & $0.2-i0.4$ &  & \\
\hline
 & & $z_{R}=2052+ i10$ &  & \\
\hline
-1, 0 & $\Sg\rho$ & $4.2+i0.1$ & &  \\
\hline
  & & {$z_{R}=1987 + i1$} &
{$z_{R}=2145+ i58$} & {$z_{R}=2383+ i73$}  \\
\hline
-1,1&$\Sg \rho$ & $1.4+i0.0$   & $-4.3-i0.7$  & $0.4+i1.1$  \\
    &$\Sg\omega$ & $1.4+i0.0$  & $1.3-i0.4$   & $-1.4-i0.4$  \\
    &$\Sg\phi$ & $-2.1-i0.0$   & $-1.9+i0.6$  & $2.1+i0.6$  \\
\hline
 & & {$z_{R}=2214+i4$} & {$z_{R}=2305+i66$}
   & {$z_{R}=2522 + i38$}  \\
\hline
-2, 1/2&$\X \rho$ & $1.8-i0.1$  & $-3.5-i1.7$  & $0.2+i1.0$   \\
       &$\X\omega$ & $1.7+i0.1$ & $2.0-i0.7$   &  $-0.6-i0.3$   \\
       &$\X\phi$ & $-2.5-i0.1$  &  $-3.0+i1.0$ & $0.9+i0.4$   \\
\hline
 &  &$z_{R}=2449 + i7$ & & \\
\hline
-3, 0&$\Om\omega$ & $1.6-i0.2$ & & \\
     &$\Om\phi$   & $-2.4+i0.3$ & & \\
\hline
\end{tabular}
\caption{The coupling $g_i$ of the resonance obtained dynamically to
the $\rho B$, $\omega B$ and $\phi B$ channels.} \label{tab:couple}
\end{center}
\end{table}

\subsection{$S=0,I=1/2$ channel}

A resonance is obtained at $z_{R}=1850+ i5MeV$ which couples to
$\Delta \rho$. We have in this case

 \be
 \label{eq:s0i12p}
 |\Delta \rho,\frac{1}{2},\frac{1}{2}\rangle
~=~\sqrt{\frac{1}{2}}|\Delta^{++} \rho^{-}\rangle
~-~\sqrt{\frac{1}{3}}|\Delta^{+} \rho^{0}\rangle
~-~\sqrt{\frac{1}{6}}|\Delta^{0} \rho^{+}\rangle
 \ee
and
 \be
\label{eq:s0i12m}
 |\Delta \rho,\frac{1}{2},-\frac{1}{2}\rangle
~=~\sqrt{\frac{1}{6}}|\Delta^{+} \rho^{-}\rangle
~-~\sqrt{\frac{1}{3}}|\Delta^{0} \rho^{0}\rangle
~-~\sqrt{\frac{1}{2}}|\Delta^{-} \rho^{+}\rangle,
 \ee

The coupling of the resonance to $\rho^0$ is obtained multiplying
the coupling of Table~\ref{tab:couple} by the corresponding
Clebsch-Gordan coefficient for $\Delta \rho^0$ of
Eqs.~(\ref{eq:s0i12p}, \ref{eq:s0i12m}). Then by means of
Eqs.~(\ref{eq:tgamma}, \ref{eq:Ggamma}), one obtains the decay
width. In this case since the $\Delta \rho^0$ component is the same
for $I_3=1/2$ and $I_3=-1/2$, one obtains the same radiative width
for the two channels, which is $\Gamma=0.722MeV$.

\subsection{$S=0,I=3/2$ channel}

One resonance is obtained at $z_{R}=1972+i49MeV$ which couples to
$\Delta \rho$, $\Delta \omega$ and $\Delta \phi$. The isospin states
for $\Delta \rho$ can be written as

 \be |\Delta \rho,\frac{3}{2},\frac{3}{2}\rangle
~=~\sqrt{\frac{3}{5}}|\Delta^{++} \rho^{0}\rangle
~+~\sqrt{\frac{2}{5}}|\Delta^{+} \rho^{+}\rangle, \ee

 \be |\Delta \rho,\frac{3}{2},\frac{1}{2}\rangle
~=~\sqrt{\frac{2}{5}}|\Delta^{++} \rho^{-}\rangle
~+~\sqrt{\frac{1}{15}}|\Delta^{+} \rho^{0}\rangle
~+~\sqrt{\frac{8}{15}}|\Delta^{0} \rho^{+}\rangle,
 \ee

\be |\Delta \rho,\frac{3}{2},-\frac{1}{2}\rangle
~=~\sqrt{\frac{8}{15}}|\Delta^{+} \rho^{-}\rangle
~-~\sqrt{\frac{1}{15}}|\Delta^{0} \rho^{0}\rangle
~+~\sqrt{\frac{2}{5}}|\Delta^{-} \rho^{+}\rangle,
 \ee

 \be |\Delta \rho,\frac{3}{2},-\frac{3}{2}\rangle
~=~\sqrt{\frac{2}{5}}|\Delta^{0} \rho^{-}\rangle
~-~\sqrt{\frac{3}{5}}|\Delta^{-} \rho^{0}\rangle. \ee

Since all the Clebsch-Gordan coefficients to $\Delta \rho^0$ are now
different, we obtain different radiative decay width for each charge
of the state. The results are $\Gamma=1.402MeV$ for $I_3=3/2$,
$\Gamma=0.143MeV$ for $I_3=1/2$, $\Gamma=0.203MeV$ for $I_3=-1/2$
and $\Gamma=1.582MeV$ for $I_3=-3/2$. It is quite interesting to see
that there is an order of magnitude difference between for $I=3/2$
and $I=1/2$, and it is a clear prediction that could be tested
experimentally.

\subsection{$S=-1,I=0$ channel}

We get a resonance at $z_{R}=2052+ i10MeV$, which couples to
$\Sigma^* \rho$. In this case

 \be |\Sigma^* \rho,0,0\rangle
~=~\sqrt{\frac{1}{3}}|{\Sigma^*}^{+} \rho^{-}\rangle
~-~\sqrt{\frac{1}{3}}|{\Sigma^*}^{0} \rho^{0}\rangle
~-~\sqrt{\frac{1}{3}}|{\Sigma^*}^{-} \rho^{+}\rangle,
 \ee
and the radiative decay obtained is $\Gamma~=~0.583MeV$.

\subsection{$S=-1,I=1$ channel}

Here we find three resonances at $z_{R}=1987 + i1MeV$, $2145+
i58MeV$ and $2383+ i73MeV$, which couple to $\Sigma^* \rho$,
$\Sigma^* \omega$ and $\Sigma^* \phi$. The relevant isospin states
are

 \be |\Sigma^* \rho,1,1\rangle
~=~\sqrt{\frac{1}{2}}|{\Sigma^*}^{+} \rho^{0}\rangle
~+~\sqrt{\frac{1}{2}}|{\Sigma^*}^{0} \rho^{+}\rangle,
 \ee

 \be |\Sigma^* \rho,1,0\rangle
~=~\sqrt{\frac{1}{2}}|{\Sigma^*}^{+} \rho^{-}\rangle
~+~\sqrt{\frac{1}{2}}|{\Sigma^*}^{-} \rho^{+}\rangle,
 \ee
and
 \be |\Sigma^* \rho,1,-1\rangle
~=~\sqrt{\frac{1}{2}}|{\Sigma^*}^{0} \rho^{-}\rangle
~-~\sqrt{\frac{1}{2}}|{\Sigma^*}^{-} \rho^{0}\rangle.
 \ee
The results obtained in this case are summarized in
Table~\ref{tab:decay-sm1i1}.

\begin{table}[hbt]
\begin{center}
\begin{tabular}{c|c|c|c}
\hline
   $I_3$           & $(1987)$ & $(2145)$  &  $(2383)$  \\
\hline
$1$              & $0.561$     & $0.399$      & $0.182$ \\
$0$              & $0.199$     & $0.206$      & $0.277$ \\
$-1$             & $0.020$     & $2.029$      & $0.537$  \\
\hline
\end{tabular}
\caption{The radiative decay widths in units of $MeV$ for the $S=-1,
I=1$ resonances with different isospin projection $I_3$.}
 \label{tab:decay-sm1i1}
\end{center}
\end{table}

\subsection{$S=-2,I=\frac{1}{2}$ channel}

Here we also find three states at $z_{R}=2214 + i4MeV$, $2305+
i66MeV$ and $2522+ i38MeV$, which couple to $\Xi^* \rho$, $\Xi^*
\omega$ and $\Xi^* \phi$. The isospin states for $\Xi^* \rho$ are
written as

\be |\Xi^* \rho,\frac{1}{2},\frac{1}{2}\rangle
~=~\sqrt{\frac{2}{3}}|{\Xi^*}^{-} \rho^{+}\rangle
~+~\sqrt{\frac{1}{3}}|{\Xi^*}^{0} \rho^{0}\rangle,
 \ee

\be |\Xi^* \rho,\frac{1}{2},-\frac{1}{2}\rangle
~=~-\sqrt{\frac{1}{3}}|{\Xi^*}^{-} \rho^{0}\rangle
~+~\sqrt{\frac{2}{3}}|{\Xi^*}^{0} \rho^{-}\rangle,
 \ee

The radiative decay widths in this case are shown in Table
~\ref{tab:decay-sm2i12}.

\begin{table}[hbt]
\begin{center}
\begin{tabular}{c|c|c|c}
\hline
   $I_3$           & $(2214)$    & $(2305)$  & $(2522)$  \\
\hline
$1/2$              & $0.815$     & $0.320$   & $0.044$  \\
$-1/2$             & $0.054$     & $1.902$   & $0.165$  \\
\hline
\end{tabular}
\caption{The radiative decay widths in units of $MeV$ for the $S=-2,
I=1/2$ resonances with the different isospin projection $I_3$.}
 \label{tab:decay-sm2i12}
\end{center}
\end{table}

\subsection{$S=-3,I=0$ channel}

Here we have only one state at $z_R=2449+i7MeV$, which couples to
$\Omega \omega$ and $\Omega \phi$. The radiative decay width
obtained in this case is $\Gamma=0.330MeV$.

As one can see, there is a large variation in the radiative width of
the different states, which should constitute a good test for the
model when these widths are measured.

In Table~\ref{tab:pdg} we summarize all the results obtained making
an association of our states to some resonances found in the PDG\cite{pdg}.

\begin{table*}[!ht]
      \renewcommand{\arraystretch}{1.5}
     \setlength{\tabcolsep}{0.2cm}
\begin{center}
\begin{tabular}{c|c|lc|l|c|c|c|c|c|c}\hline\hline
$S,\,I$& Theory & \multicolumn{2}{c|}{PDG data}
& \multicolumn{7}{c}{Predicted width $(KeV)$ for $I_3$}\\
\hline
        & pole position  & name & $J^P$ &$-3/2$ &$-1$ &$-1/2$ &$0$ &$1/2$ &$1$ & $3/2$ \\
        &$(MeV)$&&&&&&&&& \\
\hline
$0,1/2$ & $1850+i5$     & $N(2090)$ & $1/2^-$ &&&722&&722&&\\
        &                  & $N(2080)$ & $3/2^-$ &&&&&&&\\
\hline
$0,3/2$ & $1972+i49$    & $\De(1900)$ & $1/2^-$  &1582&&203&&143&&1402\\
  		&               & $\De(1940)$ & $3/2^-$  &&&&&&& \\
        &               & $\De(1930)$ & $5/2^-$ &&&&&&&   \\
\hline
$-1,0$  & $2052+i10$    & $\Lambda(2000)$ & $?^?$ &&&&583&&&\\
\hline 
$-1,1$  & $1987+i1$     & $\Sigma(1940)$ & $3/2^-$ &&20&&199&&561&  \\
        &    			& $\Sigma(2000)$ & $1/2^-$ &&2029&&206&&399& \\
    	& $2145+i58$  	& $\Sigma(2250)$ & $?^?$ &&537&&277&&182& \\
    	& $2383+i73$	& $\Sigma(2455)$ & $?^?$ &&&&&&& \\
\hline
$-2,1/2$ & $2214+i4$   & $\Xi(2250)$ & $?^?$ &&&54&&815&&\\
     & $2305+i66$      & $\Xi(2370)$ & $?^?$  &&&1902&&320&&\\
         & $2522+i38$  & $\Xi(2500)$ & $?^?$ &&&165&&44&&\\
\hline
$-3,1$   & $2449+i7$    & $\Omega(2470)$   & $?^?$  &&&&330&&& \\
 \hline\hline
    \end{tabular}
\caption{The predicted radiative decay widths of the 10 dynamically
generated resonances for different isospin projection $I_3$. Their
possible PDG counterparts are also listed. Note that the $\Sigma(2000)$ could be
the spin parter of the $\Sigma(1940)$, in which case the radiative decay widths would be
those of the $\Sigma(1940)$.}
\label{tab:pdg}
\end{center}
\end{table*}

\section{Extension to the dynamically generated states from vector meson - baryon octet interaction}
In this section we take the states dynamically generated in \cite{angelsvec} from the interacion of
vector mesons and baryons of the octet.
The generalization of the equations is rather obvious: eq. (\ref{eq:tsmvb}) becomes now ($3/2\rightarrow1/2$)
\begin{eqnarray}
\label{eq:tsmvboct}
 t_{SM\rightarrow V^\prime B^\prime}
 &=&g_i C(1/2, 1, S; m_s, M-m_s, M)\langle
 \vec{\epsilon}~|\vec{\epsilon}_{M-m_s}\rangle.
\end{eqnarray}
and equations (\ref{eq:tgamma}) and (\ref{eq:Ggamma}), which determine the radiative decay width, are identical. Once again one has to obtain
the projection of the coupling from the isospin basis to the $\rho^0 N$ case, which we detail below.

\section{Results for radiative decays into $\gamma$ and baryon octet}

The couplings of the resonances to the different $VB$ channels are
given in Ref.~\cite{sourav} in the isospin basis. For the case of
$\omega B$ and $\phi B$, there is no change to be done, but for the
case of $\rho B$, one must project over the $\rho^0 B$ component.
Since this depends on the charge of the resonance $R$, the radiative
decays will depend on this charge, as we will see. We recall that in
our phase convention $|\rho^+~\rangle=~- |1,1\rangle$ of isospin.
The information on the resonances and their couplings to different
baryons of octet and vector mesons $\rho$,$\omega$, $\phi$ for
different channels is listed in Table~\ref{tab:couple2}. We detail
the results below.

\begin{table}[hbt]
\begin{center}
\begin{tabular}{c|c|c|c|c}
\hline
S, I&Channel &  &  & \\
\hline
 & &  $z_{R}=1696$ & $z_{R}=1977+ i53$ & \\
\hline
0, 1/2&$\rho N$   & $3.2+i0$ & $-0.3-i0.5$ &  \\
      &$\omega N$ & $0.1+i0$ & $-1.1-i0.4$ &  \\
      &$\phi N$   & $-0.2+i0$ &$1.5+i0.6$  &  \\
\hline
 & & $z_{R}=1784+ i4$ & $z_{R}=1906+ i70$ & $z_{R}=2158+ i13$ \\
\hline
-1, 0 & $\omega \Lambda$ & $1.4+i0.03$  & $0.4+i0.2$ & $-0.3-i0.2$ \\
      & $\rho   \Sigma $ & $-1.5+i0.03$ & $3.1+i0.7$ & $0.01-i0.08$ \\
      & $\phi   \Lambda$ & $-1.9-i0.04$ & $-0.6-i0.3$ &$0.5+i0.3$  \\
\hline
  & & {$z_{R}=1830 + i40$} &   {$z_{R}=1987+ i240$}  \\
\hline
-1,1&$\rho \Lambda $ & $-1.6+i0.2$   & $-0.3+i0.9$    \\
    &$\rho \Sigma  $ & $-1.6+i0.07$  & $2.6+i0.0$     \\
    &$\omega \Sigma$ & $-0.9+i0.1$   & $-0.2+i0.5$    \\
    &$\phi   \Sigma$ & $1.2-i0.2$   & $0.2-i0.7$    \\
\hline
 & & {$z_{R}=2039+i67$} & {$z_{R}=2082+i31$}
     \\
\hline
-2, 1/2&$\rho   \Xi$ & $2.4+i0.7$  & $0.4+i0.3$       \\
       &$\omega \Xi$ & $0.6-i0.08$ & $1.1+i0.3$       \\
       &$\phi   \Xi$ & $-0.8+i0.1$  &  $-1.6-i0.4$    \\
\hline
\end{tabular}
\caption{The coupling $g_i$ of the resonance obtained dynamically to
the $\rho B$, $\omega B$ and $\phi B$ channels.} \label{tab:couple2}
\end{center}
\end{table}
\subsection{$S=0,I=1/2$ channel}

Two resonances are obtained at $z_{R}=1696MeV$ and $z_{R}=1977+
i53MeV$ which couple to $\rho N$, $\omega N$ and $\phi N$. We have
in this case

 \be
 \label{eq:s0i12poct}
 |\rho N,\frac{1}{2},\frac{1}{2}\rangle
~=~-\sqrt{\frac{1}{3}}|\rho^{0} p\rangle ~-~\sqrt{\frac{2}{3}}|
\rho^{+} n\rangle
 \ee
and
 \be
 \label{eq:s0i12moct}
 |\rho N,\frac{1}{2},-\frac{1}{2}\rangle
~=~\sqrt{\frac{1}{3}}|\rho^{0} n\rangle ~-~\sqrt{\frac{2}{3}}|
\rho^{-} p\rangle
 \ee

The coupling of the resonance to $\rho^0$ is obtained multiplying
the coupling of Table~\ref{tab:couple2} by the corresponding
Clebsch-Gordan coefficient for $ \rho^0 N$ of Eqs.~(\ref{eq:s0i12poct},
\ref{eq:s0i12moct}). Then by means of Eqs.~(\ref{eq:tgamma},
\ref{eq:Ggamma}), one obtains the decay width.

\subsection{$S=-1,I=0$ channel}

We get three resonances at $z_{R}=1784+ i4MeV$, $z_{R}=1906+ i70MeV$
and $z_{R}=2158+ i13MeV$ respectively, which couple to $\rho
\Sigma$,$\omega \Lambda$ and $\phi \Lambda$. In this case

 \be |\rho \Sigma,0,0\rangle
~=~\sqrt{\frac{1}{3}}| \rho^{-} \Sigma^{+}\rangle
~-~\sqrt{\frac{1}{3}}| \rho^{0} \Sigma^{0}\rangle
~-~\sqrt{\frac{1}{3}}| \rho^{+} \Sigma^{-}\rangle.
 \ee

\subsection{$S=-1,I=1$ channel}

Here we find two resonances at $1830+ i40MeV$ and $1987+ i240MeV$,
which couple to $\rho \Lambda$ $\rho \Sigma$, $\omega \Sigma$ and
$\phi \Sigma$. The relevant isospin states are

 \be |\rho \Sigma,1,1\rangle
~=~-~\sqrt{\frac{1}{2}}| \rho^{0} {\Sigma}^{+}\rangle
~-~\sqrt{\frac{1}{2}}| \rho^{+} {\Sigma}^{0}\rangle,
 \ee

 \be |\rho \Sigma,1,0\rangle
~=~-~\sqrt{\frac{1}{2}}| \rho^{+} {\Sigma}^{-}\rangle
~-~\sqrt{\frac{1}{2}}| \rho^{-} {\Sigma}^{+}\rangle
 \ee

and
 \be |\rho \Sigma,1,-1\rangle
~=~-\sqrt{\frac{1}{2}}|\rho^{-} {\Sigma}^{0}\rangle
~+~\sqrt{\frac{1}{2}}|\rho^{0} {\Sigma}^{-} \rangle.
 \ee

\subsection{$S=-2,I=\frac{1}{2}$ channel}

Here we also find two states at $z_{R}=2039 + i67MeV$ and $2082+
i31MeV$, which couple to $\rho \Xi$, $\omega \Xi$ and $\phi \Xi$.
The isospin states for $\rho \Xi$ are written as

\be |\rho \Xi,\frac{1}{2},\frac{1}{2}\rangle ~=~-\sqrt{\frac{2}{3}}|
\rho^{+} {\Xi}^{-}\rangle ~-~\sqrt{\frac{1}{3}}| \rho^{0}
{\Xi}^{0}\rangle,
 \ee

\be |\rho \Xi,\frac{1}{2},-\frac{1}{2}\rangle ~=~\sqrt{\frac{1}{3}}|
\rho^{0} {\Xi}^{-}\rangle ~-~\sqrt{\frac{2}{3}} | \rho^{-}
{\Xi}^{0}\rangle.
 \ee

In Table~\ref{tab:pdg2} we summarize all the results obtained, making
an association of our states to some resonances found in the
PDG\cite{pdg}.

\begin{table*}[!ht]
      \renewcommand{\arraystretch}{1.5}
     \setlength{\tabcolsep}{0.2cm}
\begin{center}
\begin{tabular}{c|c|lc|l|c|c|c|c}\hline\hline
$S,\,I$& Theory & \multicolumn{2}{c|}{PDG data}
& \multicolumn{5}{c}{Predicted width $(KeV)$ for $I_3$}\\
\hline
        & pole position  & name & $J^P$ & $-1$ &$-1/2$ &$0$ &$1/2$ &$1$  \\
        &$(MeV)$&&&&&&& \\
\hline
$0,1/2$ & $1696$        & $N(1650)$ & $1/2^-$ & &334 &  &253 &\\
        &                  & $N(1700)$ & $3/2^-$ &&&&&\\
        & $1977+i53$       & $N(2080)$ & $3/2^-$ &&196&&79&\\
        &                  & $N(2090)$ & $1/2^-$ &&&&&\\
\hline
$-1,0$  & $1784+i4$    & $\Lambda(1690)$ & $3/2^-$ &&&65~(166)&&\\
        & $ $    & $\Lambda(1800)$ & $1/2^-$ &&&&&\\
        & $1907+i70$    & $\Lambda(2000)$ & $?^?$ &&&321~(21)&&\\
        & $2158+i13$    &                 &       &&&0~(17)&&\\

\hline $-1,1$  & $1830+i40$   & $\Sigma(1750)$ & $1/2^-$ &363&&69~(240)&&7  \\
               & $1987+i240$  & $\Sigma(1940)$ & $3/2^-$ &307&&27~(90)&&426 \\
               &              & $\Sigma(2000)$ & $1/2^-$ &&&&& \\
\hline
$-2,1/2$ & $2039+i67$   & $\Xi(1950)$ & $?^?$ &&400&&89&\\
         & $2082+i31$  & $\Xi(2120)$ & $?^?$  &&212&&84&\\
 \hline\hline
    \end{tabular}
\caption{The predicted radiative decay widths of the nine
dynamically generated resonances for different isospin projection
$I_3$. Their possible PDG counterparts are also listed. The values
in the bracket for $I_3=0$ denote widths for the radiative decay
into $\Lambda \gamma$, while the values outside the bracket denote
widths for $\Sigma \gamma$.} \label{tab:pdg2}
\end{center}
\end{table*}

As one can see, there is a large variation in the radiative width of
the different states, which should constitute a good test for the
model. For the case of the vector-baryon octet states which decay into 
$\gamma$ and a baryon of the octet, it is customary to express the experimental 
information in terms of helicity amplitudes $A_{1/2}$ and $A_{3/2}$. 
We evaluate these amplitudes below to facilitate the comparison with experiment.

\section{Helicity amplitudes}

Recalling eq. (\ref{eq:tsmvboct}) for the dynamically generated states from a vector and a baryon of the 
octet, we have the two cases $J^P=1/2^{-}$ and $J^P=3/2^{-}$. The helicity
amplitudes are defined as 
\begin{eqnarray}
\label{eq:hadef}
 A^{N^{*}}_{1/2}&=&\sqrt{\frac{2\pi \alpha}{k}}\frac{1}{e} \left\langle N^{*},J_z=1/2|\epsilon ^{(+)}_\mu J^{\mu}|N,J_z=-1/2\right\rangle \\
 A^{N^{*}}_{3/2}&=&\sqrt{\frac{2\pi \alpha}{k}}\frac{1}{e} \left\langle N^{*},J_z=3/2|\epsilon ^{(+)}_\mu J^{\mu}|N,J_z=1/2\right\rangle
\end{eqnarray}
where $\alpha=1/137$, $k$ is the CM photon momentum and $e^2=4\pi\alpha$. To acomodate
these amplitudes to our eq. (\ref{eq:tsmvboct}) we rewrite them taking 
$\epsilon ^{(+)}_\mu J^{\mu}=-\vec{\epsilon}~^{(+)} \vec{J}$, as
\begin{equation}
\label{eq:hacgdef}
 A^{J=1/2}_{1/2} = -t_{\gamma} \frac{1}{\sqrt{2k}} C(1/2, 1, 1/2; m_s, M-m_s, M)\langle
 \vec{\epsilon}_{M-m_s} |\vec{\epsilon}^{*}\rangle ^{*}
\end{equation}
where $t_{\gamma}$ is given by eq. (\ref{eq:tgamma}), with $m_s=-1/2$, $\vec{\epsilon}=\vec{\epsilon}~^{(+)}$, which fixes $M-m_s=1$, 
and similarly for the other amplitudes. Hence, we obtain
\begin{align}
\label{eq:hacg1} 
 A^{J=1/2}_{1/2} &= -t_{\gamma} \frac{1}{\sqrt{2k}} C(1/2, 1, 1/2; -1/2, 1, 1/2) = \frac{1}{\sqrt{2k}} \sqrt{\frac{2}{3}} t_{\gamma} \\
 \label{eq:hacg2} 
 A^{J=3/2}_{1/2} &= -t_{\gamma} \frac{1}{\sqrt{2k}} C(1/2, 1, 3/2; -1/2, 1, 1/2) = -\frac{1}{\sqrt{2k}} \sqrt{\frac{1}{3}} t_{\gamma}  \\
 \label{eq:hacg3} 
 A^{J=3/2}_{3/2} &= -t_{\gamma} \frac{1}{\sqrt{2k}} C(1/2, 1, 3/2; 1/2, 1, 1/2) = -\frac{1}{\sqrt{2k}} t_{\gamma}
\end{align}
The ordinary formula to get the radiative decay width in terms of $A_{1/2}$ and $A_{3/2}$
is given in the PDG \cite{pdg} as
\begin{align}
\label{eq:dwha}
\Gamma_{\gamma} &= \frac{k^2}{\pi} \frac{2 M_B}{(2J_R+1)M_R} \left[ (A_{1/2})^2+(A_{3/2})^2 \right] 
\end{align}
One can see that using in eq. (\ref{eq:dwha}), the values of the helicity amplitudes obtained in eqs. (\ref{eq:hacg1}, \ref{eq:hacg2}, \ref{eq:hacg3}) one obtains the same result of eq. (\ref{eq:Ggamma})
 for both spins of the resonances.

It is interesting to note that the values of $A_{1/2}$ for $J=1/2,3/2$ and $A_{3/2}$ for $J=3/2$
are all related by the simple relations of eqs. (\ref{eq:hacg1}, \ref{eq:hacg2}, \ref{eq:hacg3}) for these dynamically generated
states, something that could be contrasted with experiment. We compile in Table~\ref{tab:hares} all the results obtained for the resonances that are likely to be associated to states in the PDG
for which there are data.
\begin{table*}[!ht]
      \renewcommand{\arraystretch}{1.5}
     \setlength{\tabcolsep}{0.2cm}
\begin{center}
\begin{small}
\begin{tabular}{lc|c|ccccccc}\hline\hline
 \multicolumn{2}{c|}{PDG data}
& \multicolumn{8}{c}{Helicity amplitudes $10^{-3}(GeV^{-1/2})$}\\
\hline
 name	& $J^P$			& Decay 		&	Theory	& Exp.\cite{pdg} & Exp.\cite{Barbour:1978cx} &Exp.\cite{Devenish:1974zh}		&Th.\cite{Merten:2002nz}	&Th.\cite{Carlson:1998si} &Th.\cite{Capstick:1992uc}\\
 		&				&				&			& PDG	& Barbour & Devenish & \\
\hline
 $N(1650)$ 	& $1/2^-$ 	& $A^p_{1/2}$	&$64\pm7$	&$53\pm16$	&&&$5$	&$46$	&$54$\\
			& 			& $A^n_{1/2}$	&$-74\pm7$	&$-15\pm4$	&&&$-16$	&$-58$ 	&$-35$\\
 $N(1700)$ 	& $3/2^-$ 	& $A^p_{1/2}$	&$-46\pm5$	&$-18\pm13$	&$-33\pm21$&&$-13$ 	&$-3$ 	&$-33$\\
			& 			& $A^p_{3/2}$	&$-79\pm9$	&$-2\pm24$	&$-14\pm25$&&$-10$ 	&$15$  	&$18$\\
			& 			& $A^n_{1/2}$	&$52\pm5$	&$0\pm50$	&$50\pm42$&&$16$ 	&$14$  	&$-3$\\
			& 			& $A^n_{3/2}$	&$91\pm9$	&$-3\pm44$	&$35\pm30$&&$-42$	&$-23$  &$-30$\\
\hline
 $N(2080)$	& $3/2^-$ 	& $A^p_{1/2}$	&$-21\pm5$	&$-20\pm8$	&  &$26\pm52$ \\
			& 			& $A^p_{3/2}$	&$-36\pm8$	&$17\pm11$	&  &$128\pm57$  \\
			& 			& $A^n_{1/2}$	&$-29\pm5$	&$7\pm13$	&  &$53\pm83$  \\
			& 			& $A^n_{3/2}$	&$-50\pm8$	&$-53\pm34$	&  &$100\pm141$  \\
 $N(2090)$ 	& $1/2^-$ 	& $A^p_{1/2}$	&$30\pm6$	& & & \\
			& 			& $A^n_{1/2}$	&$41\pm6$	& & & \\
 \hline\hline
    \end{tabular}
\caption{Comparison with experiments and other theories}
\label{tab:hares}
\end{small}
\end{center}
\end{table*}
The theoretical errors have been obtained by assuming $10\%$ uncertainty in the largest
coupling of the resonance to the different channels and $15\%$ in the other ones. This
is only a rough estimate and the uncertainties can easily be double this amount.
 
By looking at the table we can see that the agreement with the data of $A^p_{1/2}$
for the $N^{*}(1650)$ is good. For the case of $A^n_{1/2}$ the results obtained
are larger than experiment but the sign is good. In the case of the $N^{*}(1700)$,
$A^p_{1/2}$ can be considered qualitatively fine within theoretical and experimental
errors, $A^p_{3/2}$ seems to be larger than experiment but one can see that individual
measurements, as the one of Barbour \cite{Barbour:1978cx} diverge appreciably from
the PDG average values. Similarly $A^n_{1/2}$ would be compatible with experiment
within errors and $A^n_{3/2}$ seems also a bit larger, but not qualitatively too
off account taken of the large experimental uncertainties. This last magnitude is very 
relevant since the predictions of the dynamically generated model have opposite sign
to all the quark model calculations mentioned in the table. Since a global sign is these nondiagonal
transitions can always appear, more relevant than the absolute sign is the relative
one to $A^n_{1/2}$ which is the same in our case and opposite in \cite{Merten:2002nz,Carlson:1998si}.
In ref. \cite{Capstick:1992uc} one has the same signs but there is one order of magnitude
difference between the two helicity amplitudes, while in our model the ratio is $\sqrt{3}$.
It is clear that precise measurements of these magnitudes are very useful to discriminate among
models and help us understand better the structure of these resonances.

The case of the $N^{*}(2080)$ and $N^{*}(2090)$ is more unclear. While for $A^p_{1/2}$ and
$A^n_{3/2}$ we would obtain apparently good agreement with experiment, this is not the case
for $A^p_{3/2}$ and $A^n_{1/2}$, although the experimental uncertainties are very large. We
also show the experimental results of Devenish \cite{Devenish:1974zh} for the resonances to
show that individual measurements are very different from the PDG averages. Since under the
umbrella of the $N^{*}(2080)$ and $N^{*}(2090)$ there are apparently different states compiled,
it would be possible that the averages of the PDG were not done for different measurements
on the same state but for measurements on different states. The experimental situation is
hence unclear but the results obtained here should be a motivation for further
reanalysis.

\begin{table*}[!ht]
      \renewcommand{\arraystretch}{1.5}
     \setlength{\tabcolsep}{0.2cm}
\begin{center}
\begin{tabular}{cc}
$N^*(1650)$ & $N^*(1700)$ \\
 \includegraphics[width=0.5\textwidth]{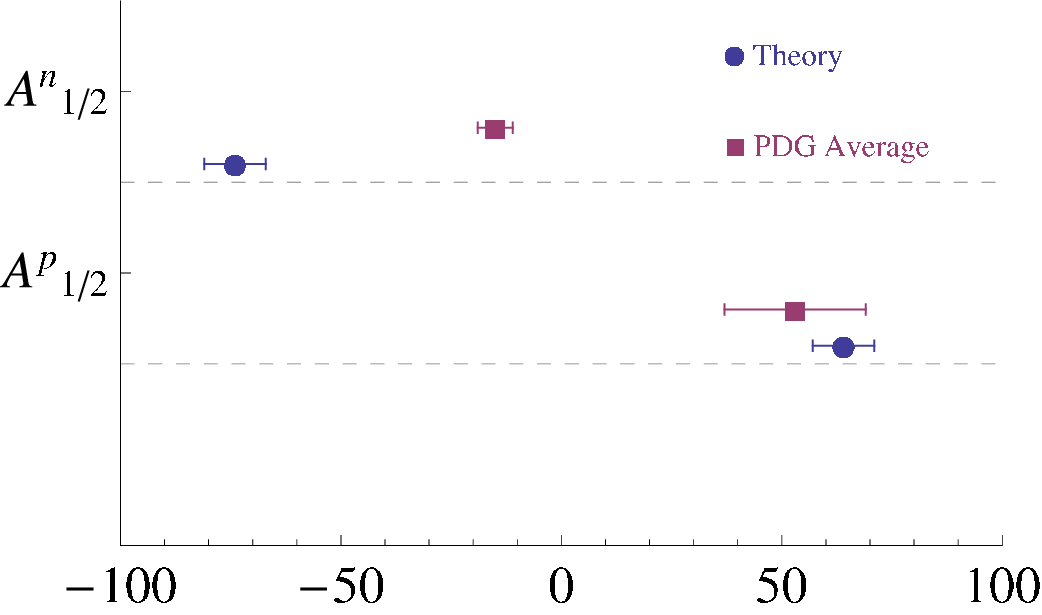} & \includegraphics[width=0.5\textwidth]{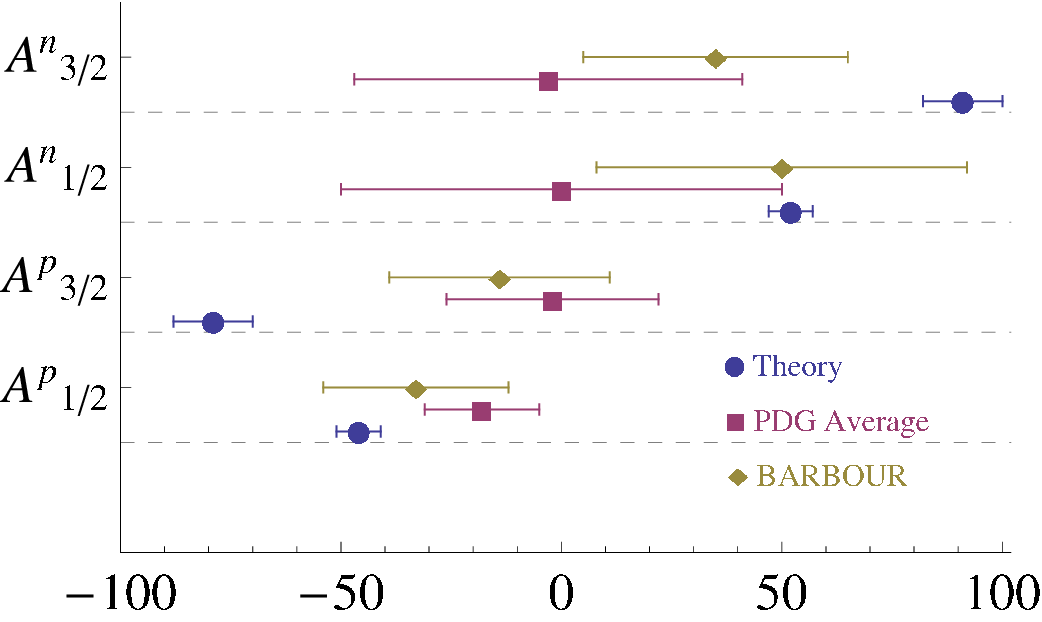} \\
 $N^*(2080)$ & $N^*(2090)$ \\
 \includegraphics[width=0.5\textwidth]{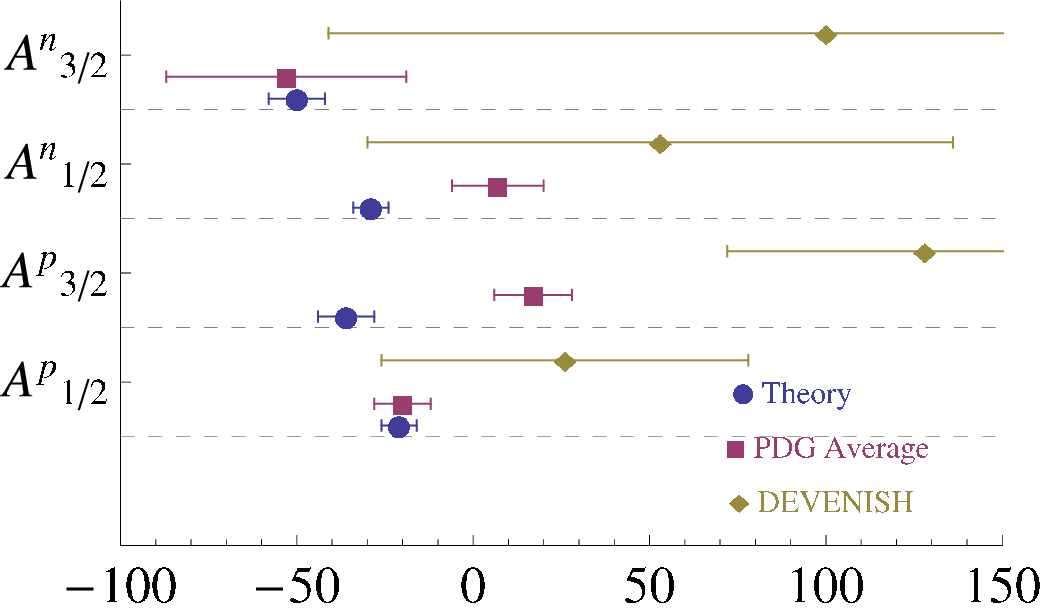} & \includegraphics[width=0.5\textwidth]{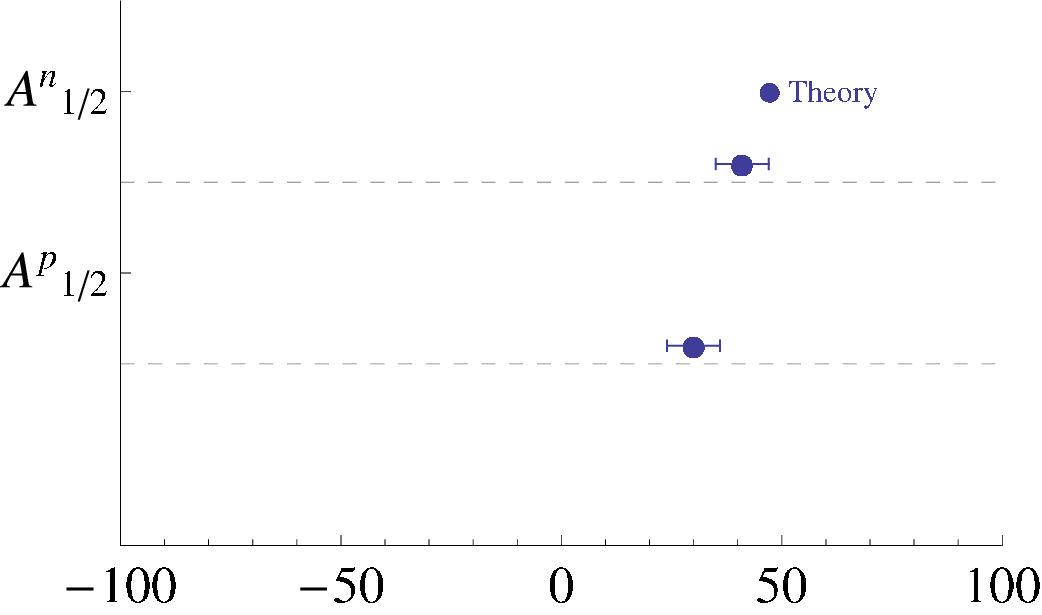} \\
    \end{tabular}
\caption{Graphical view of the experimental data with the results of the theory. The units of
$A^N_{\lambda}$ are in $10^{-3} GeV^{-1/2}$.}
\label{tab:hagraph}
\end{center}
\end{table*}
In the Table~\ref{tab:hagraph} we present the results of Table~\ref{tab:hares} 
to give a global view of how the results compare with the data. With the coments 
expressed above in the detailed discussion one could qualify as qualitatively fair the global
agreement of the results with the data.
\section{Conclusions}

We have studied the radiative decay into $\gamma B$, with $B$ a baryon of the
octet and decuplet of SU(3), of the dynamically generated resonances obtained  within
the framework of the local hidden gauge mechanism for vector interactions. The
framework is particularly rewarding for the study of such observable, since the
photon in the final state appears coupling directly to the vector
 $V= \rho^0, \omega, \phi$  in the $R \to V$ amplitudes which are
 studied in previous works.  
 The rates obtained are large and the radiative widths are of the order of $1
 ~MeV$. On the other hand, one of the appealing features of the results is the
 large difference, of about one order of magnitude, that one finds between the
 widths for different charge states of the same particle. These results are tied
 to  details of the theory, concretely the coupling of the resonances to
 $ V~B$, which sometimes produce large interferences between the different
 contributions of the three vector mesons to which the photon couples.
  As a consequence, the radiative decay widths that we have evaluated bear much
  information on the nature of those resonances, which should justify efforts
  for a systematic measurement of these observables. 
  
  We have studied the decay into $\gamma$-baryon octet and  $\gamma$-baryon decuplet
  of the states dynamically generated from the vector-baryon octet and vector-baryon
  decuplet interaction. In the first case one can define the helicity amplitude $A_{1/2}$
  and $A_{3/2}$ for the $n$ and $p$ type states of the $N^{*}$, which makes the comparison
  with data more useful. We have found good agreement with data in some cases and rough
  in others, but we have warned about the large experimental uncertainties and the 
  possibility that the PDG averages are done over different states. What stands clear
  from the work and the discussion is that these observables are very useful to help
  us understand better the nature of the resonances discussed here. Further experimental
  work is most desirable. We hope the present work stimulates work in this direction.

\section*{Acknowledgments}

We would like to thank M. J. Vicente Vacas for a critical reading of the
manuscript. This work is partly supported by DGICYT contract number
FIS2006-03438.
This research is  part of the EU Integrated Infrastructure Initiative Hadron
 Physics Project
under  contract number RII3-CT-2004-506078. B. X. Sun acknowledges support
from the National Natural Science Foundation of China under grant number
10775012.

\end{document}